\newcommand{\be}{\begin{equation}}
\newcommand{\ee}{\end{equation}}
\newcommand{\bea}{\begin{eqnarray}}
\newcommand{\eea}{\end{eqnarray}}

\newcommand{\Appendix}[1]%
    {\renewcommand{\thesection}{Appendix~\Alph{section}:}%
         \section{#1}}%
\newcommand*\patchAmsMathEnvironmentForLineno[1]{%
  \expandafter\let\csname old#1\expandafter\endcsname\csname #1\endcsname
  \expandafter\let\csname oldend#1\expandafter\endcsname\csname end#1\endcsname
  \renewenvironment{#1}%
     {\linenomath\csname old#1\endcsname}%
     {\csname oldend#1\endcsname\endlinenomath}}%
\newcommand*\patchBothAmsMathEnvironmentsForLineno[1]{%
  \patchAmsMathEnvironmentForLineno{#1}%
  \patchAmsMathEnvironmentForLineno{#1*}}%
\AtBeginDocument{%
\patchBothAmsMathEnvironmentsForLineno{equation}%
\patchBothAmsMathEnvironmentsForLineno{align}%
\patchBothAmsMathEnvironmentsForLineno{flalign}%
\patchBothAmsMathEnvironmentsForLineno{alignat}%
\patchBothAmsMathEnvironmentsForLineno{gather}%
\patchBothAmsMathEnvironmentsForLineno{multline}%
}

\catcode`@=11
\long\def\@makecaption#1#2{
   \vskip 10pt
   \setbox\@tempboxa\hbox{{\small\bf #1.} \ {\small #2}}
   \ifdim \wd\@tempboxa >\hsize       
   {\small\bf #1.} \ {\small #2}\par  
   \else                              
        \hbox to\hsize{\hfil\box\@tempboxa\hfil}
   \fi}
\catcode`@=12

\catcode`@=11
\def\secteqno{\@addtoreset{equation}{section}%
\def\theequation{\thesection.\arabic{equation}}}
\def\endsecteqno{\def\theequation{\@ifundefined{chapter}%
{\arabic{equation}}{\thechapter.\arabic{equation}}}}
\newcounter{subequation}
\def\thesubequation{\alph{subequation}}
\def\sneqnarray{\stepcounter{equation}\let\@currentlabel=\theequation
\setcounter{subequation}{1}
\def\@eqnnum{{\rm (\theequation\thesubequation)}}
\global\@eqcnt\z@\tabskip\@centering\let\\=\@eqncr\let\@@eqncr=\@@sneqncr
$$\halign to \displaywidth\bgroup\@eqnsel\hskip\@centering
 $\displaystyle\tabskip\z@{##}$&\global\@eqcnt\@ne
 \hskip 2\arraycolsep \hfil${##}$\hfil
 &\global\@eqcnt\tw@ \hskip 2\arraycolsep
$\displaystyle\tabskip\z@{##}$\hfil
tabskip\@centering&\llap{##}\tabskip\z@\cr}
\def\endsneqnarray{\@@sneqncr\egroup $$\global\@ignoretrue}
\def\@@sneqncr{\let\@tempa\relax
   \ifcase\@eqcnt \def\@tempa{& & &}\or \def\@tempa{& &}
   \else \def\@tempa{&}\fi
     \@tempa \if@eqnsw\@eqnnum\stepcounter{subequation}\fi
     \global\@eqnswtrue\global\@eqcnt\z@\cr}
\def\nobiblabels{\def\@lbibitem[##1]##2{\@bibitem{##2}}}
\catcode`@=12

\def\beq{\begin{equation}}
\def\eeq{\end{equation}}
\def\bea{\begin{eqnarray}}
\def\eea{\end{eqnarray}}

\documentclass[preprint, aps, superscriptaddress, prd, showpacs, 
amsmath, amssymb, nofootinbib, showkeys]{revtex4-1}

\usepackage[german,english]{babel}
\usepackage{hyperref}
\usepackage{bm}
\usepackage{bbm}
\usepackage{braket}
\usepackage{slashed}
\usepackage{multirow}
\usepackage{epsfig}
\usepackage{epstopdf}
\usepackage{diagbox}
\usepackage[mathlines]{lineno}
\usepackage[utf8]{inputenc}
\usepackage{subfigure}
\usepackage{diagbox}
\usepackage {xcolor}

\allowdisplaybreaks

\begin{document}

\title{Unveiling Nucleon 3D Chiral-Odd Structure with Jet Axes}
\author{Wai Kin Lai}
\email{wklai@m.scnu.edu.cn}
\affiliation{Guangdong Provincial Key Laboratory of Nuclear Science, Institute of Quantum Matter, South China Normal University, Guangzhou 510006, China}
\affiliation{Guangdong-Hong Kong Joint Laboratory of Quantum Matter,
Southern Nuclear Science Computing Center, South China Normal University, Guangzhou 510006, China}
\affiliation{Department of Physics and Astronomy, University of California, Los Angeles, CA 90095, USA}
\author{Xiaohui Liu}
\email{xiliu@bnu.edu.cn}
\affiliation{Center of Advanced Quantum Studies, Department of Physics,
Beijing Normal University, Beijing 100875, China}
\affiliation{Center for High Energy Physics, Peking University, Beijing 100871, China}
\author{Manman Wang}
\affiliation{Center of Advanced Quantum Studies, Department of Physics,
Beijing Normal University, Beijing 100875, China}
\author{Hongxi Xing}
\email{hxing@m.scnu.edu.cn}
\affiliation{Guangdong Provincial Key Laboratory of Nuclear Science, Institute of Quantum Matter, South China Normal University, Guangzhou 510006, China}
\affiliation{Guangdong-Hong Kong Joint Laboratory of Quantum Matter,
Southern Nuclear Science Computing Center, South China Normal University, Guangzhou 510006, China}

\date{\today}

\begin{abstract}
We reinterpret jet clustering  
as an axis-finding procedure which, along with the proton beam, defines the virtual-photon transverse momentum $q_T$ in deep inelastic scattering (DIS). In this way, we are able to probe the nucleon intrinsic structure using jet axes in a fully inclusive manner, similar to the Drell-Yan process. We present the complete list of azimuthal asymmetries and the associated factorization formulae at leading power for deep-inelastic scattering of a nucleon. The factorization formulae involve both the conventional time-reversal-even (T-even) jet function and the T-odd one, which have access to all transverse-momentum-dependent parton distribution functions (TMD PDFs) at leading twist.
Since the factorization holds as long as $q_T \ll Q$, where $Q$ is the photon virtuality, the jet-axis probe into the nucleon structure should be feasible for machines with relatively low energies such as the Electron-Ion Collider in China (EicC). We show that, within the winner-take-all (WTA) axis-finding scheme, the coupling between the T-odd jet function and the quark transversity or the Boer-Mulders function could induce sizable azimuthal asymmetries at the EicC, the EIC and HERA.  
We also give predictions for the azimuthal asymmetry of back-to-back dijet production in $e^+e^-$ annihilation at Belle and other energies.

\end{abstract}

\maketitle

\section{Introduction}
Recently jets and jet substructure have been proposed as alternative probes for
portraying the full three-dimensional (3D) image of a nucleon and enriched the content of the transverse-momentum-dependent (TMD) spin physics~\cite{Kang:2017glf,Liu:2018trl,Gutierrez-Reyes:2019msa,Arratia:2020nxw,Liu:2020dct,Gutierrez-Reyes:2018qez,Gutierrez-Reyes:2019vbx,Kang:2022dpx,Kang:2020xyq}. The jet probe into the nucleon structure has been shown to be able to access the TMD parton distribution functions, including the Sivers function of a transversely polarized nucleon. Conventionally, we require the jets to acquire large transverse momenta, and therefore jets are regarded only feasible for high-energy colliders such as the LHC but practically challenging for machines with a relatively low center-of-mass energy such as the Electron-Ion Collider in China (EicC)~\cite{Anderle:2021wcy} or detectors more optimized for low energy scales, such as the EIC Comprehensive Chromodynamics Experiment (ECCE)~\cite{AbdulKhalek:2021gbh}. However, in this work, we will argue that this is not the case by reinterpreting jet clustering as an axis-finding procedure to measure the virtual photon $q_T$, which allows an inclusive probe of the TMD spin physics suitable also for low energy machines~\cite{liutalk}. 

In order to maximize the full outreach of the jet probe into the complete list of the nucleon spin structure, the concept of the time-reversal-odd (T-odd) jet function was proposed recently~\cite{Liu:2021ewb}. The T-odd jet function couples directly to the chiral-odd nucleon parton distributions, such as the quark transversity and the Boer-Mulders function of the proton. It immediately opens up many unique opportunities for probing the nucleon intrinsic spin dynamics using jets, which were thought to be impossible. 
Besides, the T-odd jet function is interesting by its own, since it could ``film" the QCD non-perturbative dynamics by continuously changing the jet axis from one to another. 

In this work, we study the phenomenology of the T-odd jet function in deep-inelastic scattering (DIS) of a nucleon and $e^+e^-$ annihilation. In Section~\ref{sec:measure_qt}, we explain how the jet axis is used for measuring the photon $q_T$ in a fully inclusive way and argue why the jet-axis probe of the nucleon spin and the TMDs is feasible even for low-energy machines such as the Electron-Ion Collider in China (EicC) and Belle~\cite{Accardi:2022oog}.
In Section~\ref{sec:T_odd_jet}, we briefly review the notion of the T-odd jet function. In Section~\ref{sec:DIS}, we give the complete list of the azimuthal asymmetries in the jet-axis probe in deep-inelastic scattering of a nucleon. We give predictions on the azimuthal asymmetries associated with the couplings of the T-odd jet function with the quark transversity and the Boer-Mulders function at the EicC, the EIC~\cite{AbdulKhalek:2021gbh}, and HERA.
In Section~\ref{sec:ee}, we study the azimuthal asymmetry of back-to-back dijet production in $e^+e^-$ annihilation, which is induced by the T-odd jet function. In Section~\ref{sec:summary}, we give a summary and an outlook.

\section{Measuring photon $q_T$ in DIS}\label{sec:measure_qt}

All conventional probes of the TMDs and the spin structure are more or less equivalent to measuring the virtual-photon transverse momentum $q_T$ with respect to two pre-defined axes. For instance, in the Drell-Yan process, the incoming nucleon beams naturally set up the $\pm z$-axis and the photon transverse momentum is then straightforwardly determined. In DIS, since we only have one nucleon beam, we thus need another direction to define the photon $q_T$. Tagging a final-state hadron becomes a natural option for this purpose, and in this case, the photon $q_T$ is then measured with respect to the nucleon beam and the tagged hadron momentum $P_h$. This is nothing but the semi-inclusive deep inelastic scattering (SIDIS). 

\begin{figure}[htbp]
  \begin{center}
   \includegraphics[scale=0.35]{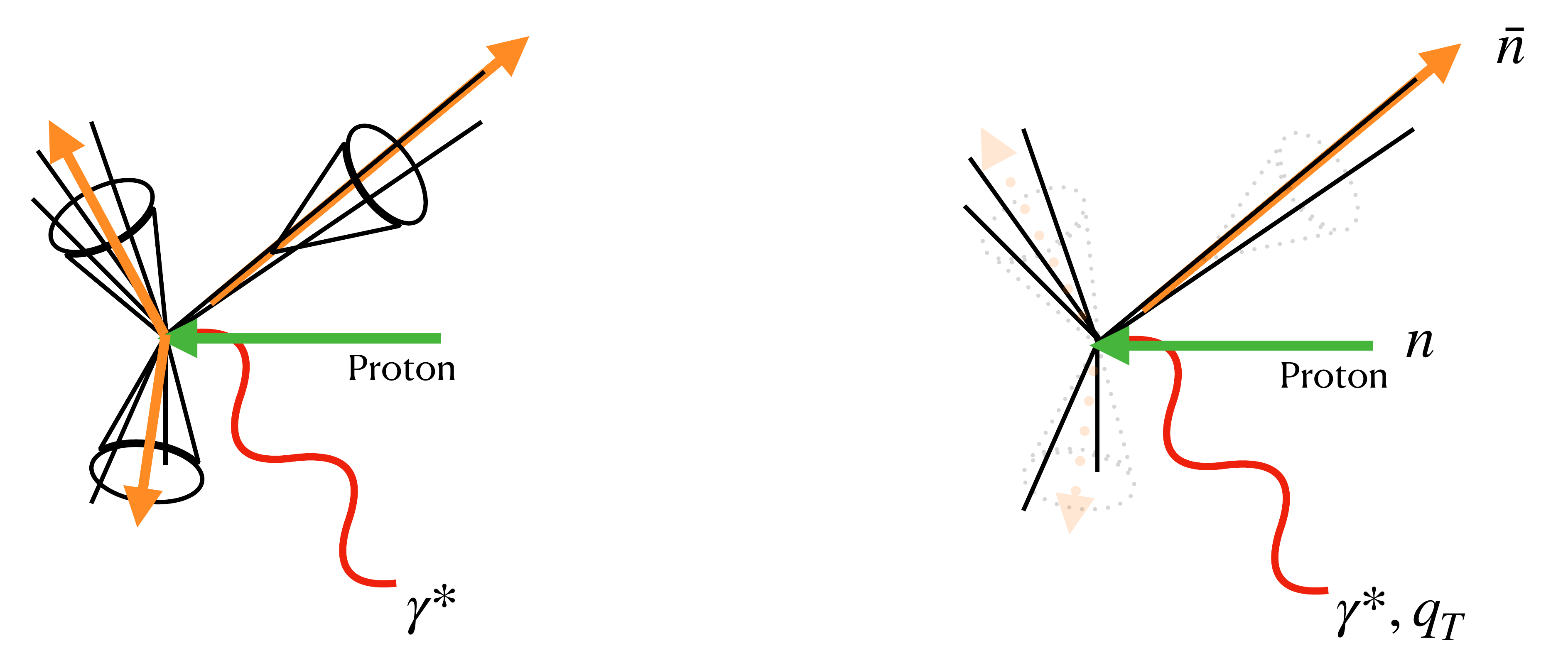} 
\caption{Photon $q_T$ by the jet axis in DIS, where one of the jet axes and the proton beam determine the ${\bar n}$ and $n$ directions respectively, which decide the photon $q_T$. }
  \label{fg:qt-jet}
 \end{center}
\end{figure}

Finding an axis for measuring the photon $q_T$ in DIS is certainly not limited to tagging hadrons. Many other strategies could also help here, such as the final-state-particle clustering. The procedure follows exactly the jet clustering algorithms, but a with different emphasis. Here the jet clustering procedure is barely a recursive algorithm for us to determine the axes, which once being determined, we measure the photon $q_T$ with respect to one of them and the proton beam to probe the nucleon structure, while totally forget about the jet, as illustrated in Fig.~\ref{fg:qt-jet}. Therefore, the jet-axis probe is fully differential just like the SIDIS. 

Based on what we have described, we can derive the factorization formula for the jet-axis probe. Formally, the factorization theorem reads
\bea 
d \sigma \propto f_i^{U/T}(x,p_T^2) \otimes {\cal J}_{1,i}(z,k_T^2) 
+ g_i^{U/T}(x,p_T^2)\otimes{\cal J}_{T,i}(z,k_T^2)  \,, \label{eq:fact_1}
\eea 
where 
$f_i^{U/T}(x,p_T^2) $ and $g_i^{U/T}(x,p_T^2)$ are the proton partonic transverse-momentum distributions for parton flavor $i$, and  
${\cal J}_{1,i}(z,k_T^2)$ and ${\cal J}_{T,i}(z,k_T^2)$ are both the jet-axis-finding functions (jet functions) which encode the perturbatively calculable jet clustering procedure. 
The conventional jet function ${\cal J}_{1,i}(z, k_T^2)$ is induced by an unpolarized quark, while a transversely polarized quark  gives rise to the time-reversal odd (T-odd) jet function ${\cal J}_{T,i}(z,k_T^2)$. The detailed factorization form and the definition of the T-odd jet function, ${\cal J}_{T,i}(z,k_T^2)$, will be given in the following sections. 
 Here we note that the factorization theorem holds as long as $Q \gg k_T $, which shares the same requirement for the SIDIS factorization to be valid. In this sense, just like the SIDIS,  the jet-axis probe will also be low-energy-machine friendly, and could likely be implemented at the EicC.
 
  To adapt the jet axis finding procedure to low-energy machines, instead of using the usual $k_T$-type jet algorithms that are widely used at the LHC, in DIS we default to the energy-type jet algorithms which is more feasible for clustering particles with low transverse momenta and populated in the forward/backward rapidities. For instance, we can adopt the spherically-invariant jet algorithm~\cite{Cacciari:2011ma}, defined by
\begin{align}
d_{ij}={\rm min}(E^{-2}_i,E^{-2}_j)\frac{1-\cos\theta_{ij}}{1-\cos R}\,,\quad d_{iB}=E_i^{-2}\,.\label{eq:spherical_jet_alg}
\end{align}
where $\theta_{ij}$ is the angle between particles $i$ and $j$, while $E_{i}$ and $E_j$ are the energy carried by them. For TMD studies, the radius parameter $R$ will be chosen such that $R \sim {\cal O}(1) \gg q_T/Q$.

\section{T-odd jet function} \label{sec:T_odd_jet}
The inclusive photon $q_T$ cross sections with respect to the proton beam and the jet axis can be written in terms of a factorization theorem Eq.~(\ref{eq:fact_1}) derived from the soft-collinear effective theory 
(SCET)~\cite{Bauer:2000yr, Bauer:2001ct, Bauer:2001yt, Bauer:2002nz}. The factorization theorem involves the 
transverse-momentum-dependent (TMD) correlator  
\begin{align}
{\cal J}^{ij}(z,k_T)&=\frac{1}{2z}\sum_X\int \frac{dy^+d^2\bm{y}_T}{(2\pi)^3}e^{ik\cdot y}
\langle 0|{\chi}^i_{\bar{n}}(y)|JX\rangle\langle JX|\bar{\chi}^j_{\bar{n}}(0)|0\rangle |_{y^-=0}\,, \label{eq:jet_corr}
\end{align}
where $\bar{n}$ is a light-like vector along the direction of the jet, $\chi_{\bar{n}}=W^\dagger_{\bar{n}}\xi_{\bar{n}} $ is the product of the collinear quark field $\xi_{\bar{n}}$ and the collinear Wilson line  $W^\dagger_{\bar{n}}$. Here, $z$ is the momentum fraction of the jet with respect to the fragmenting parton which initiates the jet, i.e. $z=P_J^-/ k^-$, with $P_J$ being the jet momentum that defines the jet axis, and $k$ the momentum of the fragmenting quark.
 The jet algorithm dependence is implicit in Eq.~(\ref{eq:jet_corr}), which determines the $P_J$ and hence the jet axis, and can be calculated perturbatively. 

Conventionally, only the chiral-even Dirac structure $\slashed{\bar{n}}$ in Eq.~(\ref{eq:jet_corr}) was considered. However, as noted in Ref.~\cite{Liu:2021ewb}, in the nonperturbative regime in which $k_T \sim \Lambda_{\rm QCD}$, 
spontaneous chiral symmetry breaking 
leads to a nonzero component of the jet which is both time-reversal-odd (T-odd) and chiral-odd, when the jet axis is different from the direction of the
fragmenting parton.
Therefore,
the correlator in Eq.~(\ref{eq:jet_corr}) in general is a sum of two structures:
\begin{align}
{\cal J}(z,k_T)&= {\cal J}_1(z,k_T^2)\frac{\slashed{\bar{n}}}{2}
+i{\cal J}_T(z,k_T^2)\frac{\slashed{k}_T\slashed{\bar{n}}}{2}\,,     \label{eq:jet_expd}
\end{align}
where ${\cal J}_1(z,k_T^2)$ is the traditional jet function, and ${\cal J}_T(z,k_T^2)$ is the T-odd jet function. Due to its chiral-odd nature, an immediate application of the T-odd jet function is to probe the chiral-odd TMD PDFs of the nucleons in DIS, such as the Boer-Mulder function and the transversity, which were thought to be impossible to access using jets.

The T-odd jet function has the following advantages:
\begin{itemize}
\item
{\textit{Universality}}\\
 Like the traditional jet function, the T-odd jet function is process independent.
\item
{\textit{Flexibility}}\\
The flexibility of choosing a jet recombination scheme and hence the jet axis allows us to adjust sensitivity of the jet function to different nonperturbative contributions. This provides an opportunity to “film” the QCD nonperturbative dynamics, if one continuously changes the axis from one to another.
\item
{\textit{Perturbative predictability}} \\
Since a jet contains many hadrons, the jet function has more perturbatively calculable degrees of freedom than the fragmentation function. For instance, in the winner-take-all (WTA) scheme, for $R\sim \mathcal{O}(1)\gg |\bm{q}_T|/E_J$, the $z$-dependence in the jet function is completely determined~\cite{Gutierrez-Reyes:2019vbx}:
\begin{align}
{\cal J}(z,k_T,R)=\delta(1-z) {\mathfrak J}(k_T)+ {\cal O}\left( \frac{k_T^2}{E_J^2R^2} \right)\,. \label{eq:WTA_J}
\end{align}
\item
{\textit{Nonperturbative predictability}}\\
Similar to the study in Ref.~\cite{Becher:2013iya}, the T-odd jet function can be factorized into a product of a perturbative coefficient
 and a nonperturbative factor. The nonperturbative factor has an operator definition~\cite{Vladimirov:2020umg},
and as a vacuum matrix element, it can be calculated on the lattice~\cite{Shanahan:2020zxr, Zhang:2020dbb}. This is unlike the TMD fragmentation function, which is an operator element with a final-state hadron tagged, making evaluation on the lattice impossible by known techniques.
\end{itemize}

The T-odd jet function will show up in various jet observables which are sensitive to nonperturbative physics. In the following, 
we study the azimuthal asymmetries in the jet-axis probe in DIS and back-to-back dijet production in $e^+e^-$ annihilation.

\section{Photon $q_T$ with respect to the jet axis in deep-inelastic scattering} \label{sec:DIS}

\begin{figure}[htbp]
  \begin{center}
 \subfigure[]{ 
   \includegraphics[scale=0.6]{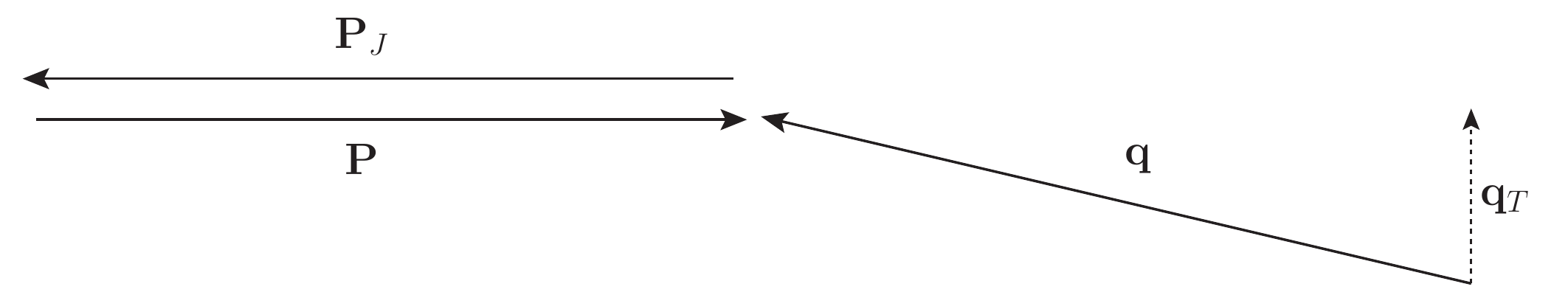}
     }
   \subfigure[]{ 
   \includegraphics[scale=0.6]{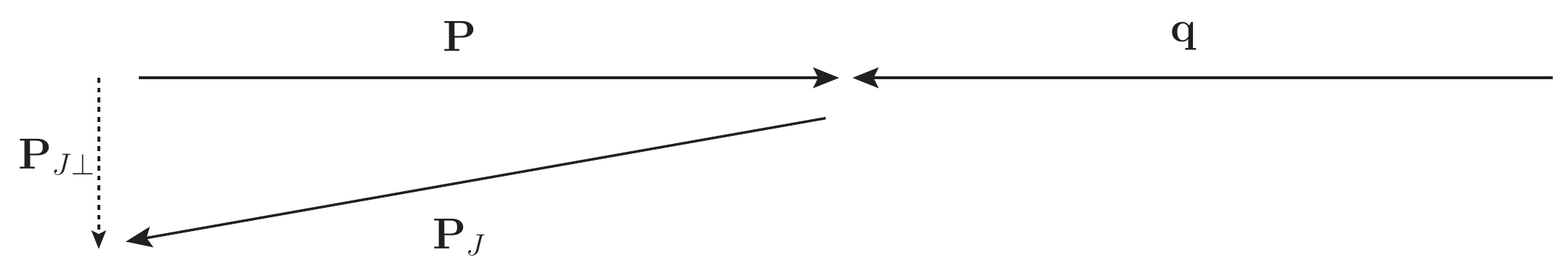}
  }
\caption{Axes in DIS in different frames: (a) the factorization frame, in which ${\bm q}_T$ is defined, and (b) the GNS system, in which ${\bm P}_{J\perp}$ is defined.}
  \label{fg:frames}
  \end{center}
\end{figure}

Consider deep-inelastic scattering of an electron off a polarized nucleon $e^-(l)+N(P)\to  e^-(l')+J(P_J)+X$, ($N=p,n$), in which we tag a jet and specify the jet axis with some recombination scheme.
We define the ${\bm q}_T$ of the virtual photon by going to the so-called factorization frame,
in which the proton beam direction and the jet axis direction are exactly opposite to each other, as shown in Fig.~\ref{fg:frames} (a). Alternatively, one can go to the gamma-nucleon system (GNS), a frame in which the virtual photon momentum and the proton beam are head-to-head (including the case of proton being at rest), and define ${\bm P}_{J\perp}$ of the jet as in Fig.~\ref{fg:frames} (b). One can show that
${\bm q}_T=-{\bm P}_{J\perp}/z$ up to corrections of order $1/Q^2$. Therefore, measuring ${\bm P}_{J\perp}$ is 
equivalent to measuring ${\bm q}_T$. In the following, we will describe the kinematics in the GNS system, which is a convention commonly used in SIDIS~\cite{Bacchetta:2006tn}.

 \begin{figure}[htbp]
  \begin{center}
   \includegraphics[scale=0.3]{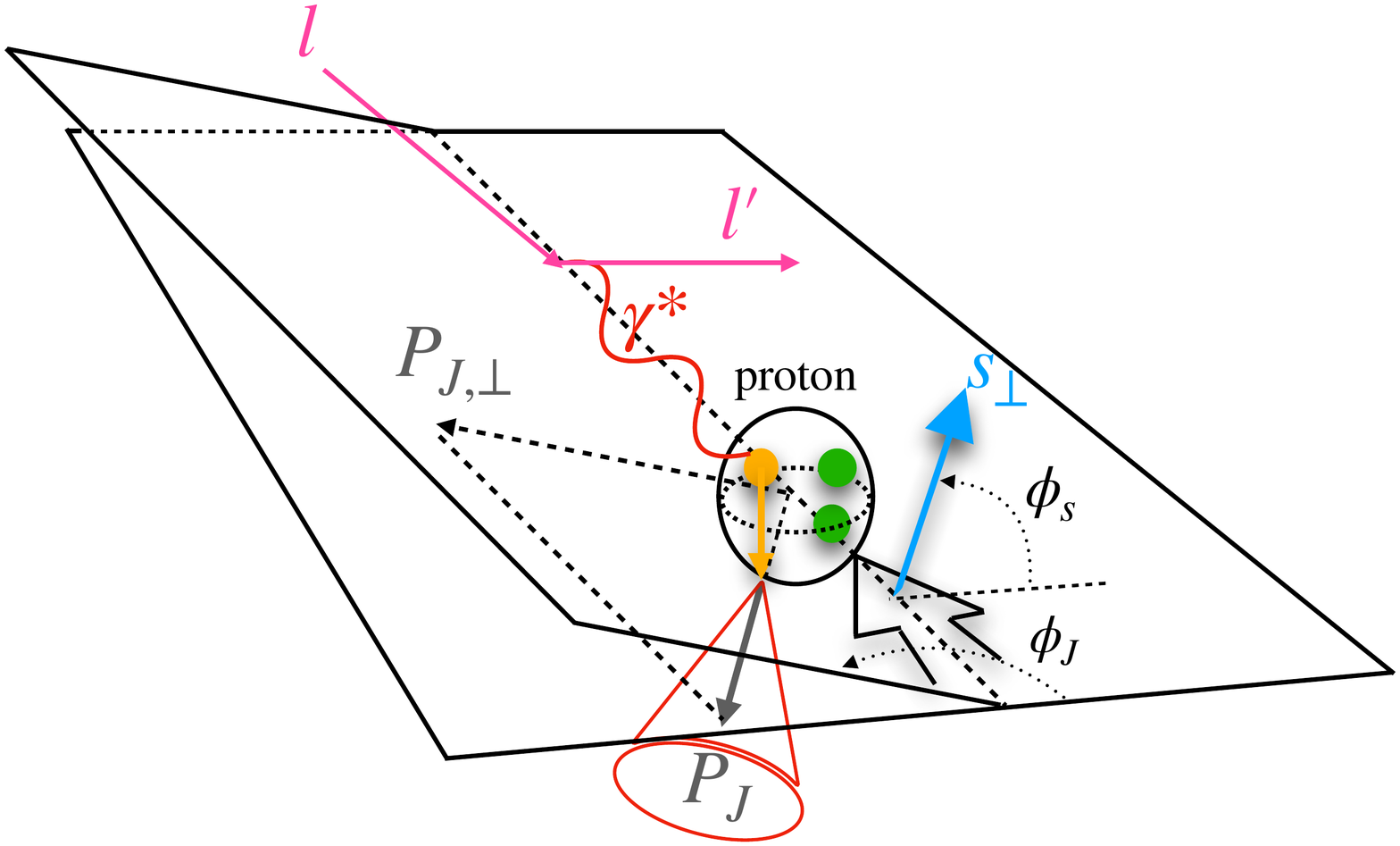} 
\caption{Kinematic configuration of DIS in the GNS system.}
  \label{fg:ep}
 \end{center}
\end{figure}

Let $M$ be the mass of the nucleon $N$ and $q=l-l'$ is the momentum carried by the virtual photon with virtuality $Q^2=-q^2$. We introduce the invariant variables 
\begin{align}
x=\frac{Q^2}{2P\cdot q}\,,\quad 
y=\frac{P\cdot q}{P\cdot l}\,,\quad
z=\frac{P\cdot P_J}{P\cdot q}\,,\quad
\gamma=\frac{2Mx}{Q}\,.
\end{align}
In the nucleon rest frame,
we can define the perpendicular component of any 3-vector as the component perpendicular to the virtual photon momentum,  ${\bm q}$. Equivalently, in Lorentz invariant notations, given any 4-vector $v^\mu$, we
define its perpendicular component by $v^\mu_\perp=g_\perp^{\mu\nu}v_\nu$, where
\begin{align}
g_\perp^{\mu\nu}&=g^{\mu\nu}-\frac{q^\mu P^\nu+P^\mu q^\nu}{P\cdot q(1+\gamma^2)}+\frac{\gamma^2}{1+\gamma^2}
\left(\frac{q^\mu q^\nu}{Q^2}-\frac{P^\mu P^\nu}{M^2}\right)\,.
\end{align} 
The 3-momenta ${\bm l}$ and ${\bm l}'$ define a plane, with respect to which we can define the azimuthal angle of any 3-vector perpendicular to ${\bm q}$. In Lorentz invariant notations, this is equivalent to defining the azimuthal angle of any 4-vector $v^\mu$ by
\begin{align}
\cos \phi_v=-\frac{l_\mu v_\nu g_\perp^{\mu\nu}}{\sqrt{l_\perp^2 v^2_\perp}}\,,\quad 
\sin\phi_v=-\frac{l_\mu v_\nu \epsilon^{\mu\nu}}{\sqrt{l_\perp^2 v^2_\perp}}\,,
\end{align}
where 
\begin{align}
\epsilon_\perp^{\mu\nu}&=\epsilon^{\mu\nu\rho\sigma}\frac{P_\rho q_\sigma}{P\cdot q\sqrt{1+\gamma^2}}\,,
\end{align}
with $\epsilon^{0123}=1$. We denote the azimthal angles of the jet momentum $P_J$ and the nucleon spin $S$ by $\phi_J$ and $\phi_S$ respectively. 
The definitions of $P_{J\perp}$, $\phi_J$, and $\phi_S$ are depicted pictorially in Fig.~\ref{fg:ep}. The nucleon spin is decomposed as sum of a longitudinal and a perpendicular component,
\begin{align}
S^\mu=S_\parallel\frac{P^\mu-\frac{M^2}{P\cdot q}q^\mu }{M\sqrt{1+\gamma^2}}+S^\mu_\perp\,,
\quad   S_\parallel=\frac{S\cdot q}{P\cdot q}\frac{M}{\sqrt{1+\gamma^2}}\,.
\end{align}
The helicity of the incoming electron is denoted by $\lambda_e$.
We define $\psi$ as the azimuthal angle of $\bm l'$ around $\bm l$.
The fully differential cross section has the most general form given by
\begin{align}
&\frac{d\sigma}{dxdyd\psi dz d\phi_J dP_{J\perp}^2}=
\frac{\alpha}{xyQ^2}\frac{y^2}{2(1-\epsilon)}\left(1+\frac{\gamma^2}{2x}\right)\left\{ F_{UU,T}+\epsilon F_{UU,L}
+\sqrt{2\epsilon(1+\epsilon)}\cos\phi_J F_{UU}^{\cos\phi_J}\right.\nonumber\\
&\quad\quad\quad\quad\quad\quad\quad\quad\quad
+\epsilon\cos(2\phi_J)F^{\cos 2\phi_J}_{UU}+\lambda_e\sqrt{2\epsilon(1-\epsilon)}\sin\phi_J F^{\sin\phi_J}_{LU}\nonumber\\
&\quad\quad\quad\quad\quad\quad\quad\quad\quad
+S_\parallel \left[\sqrt{2\epsilon(1+\epsilon)}\sin\phi_J F_{UL}^{\sin\phi_J}+\epsilon\sin(2\phi_J)F_{UL}^{\sin 2\phi_J}\right]\nonumber\\
&\quad\quad\quad\quad\quad\quad\quad\quad\quad
+S_\parallel \lambda_e\left[\sqrt{1-\epsilon^2}F_{LL}+\sqrt{2\epsilon(1-\epsilon)}\cos\phi_J F_{LL}^{\cos\phi_J}\right]\nonumber\\
&\quad\quad\quad\quad\quad\quad\quad\quad\quad +|{\bm S}_{\perp}|\left[\sin(\phi_J-\phi_S)\left(F_{UT,T}^{\sin(\phi_J-\phi_S)}+\epsilon F_{UT,L}^{\sin(\phi_J-\phi_S)}\right)\right.\nonumber\\
&\quad\quad\quad\quad\quad\quad\quad\quad\quad+\epsilon\sin(\phi_J+\phi_S)F_{UT}^{\sin(\phi_J+\phi_S)}+\epsilon\sin(3\phi_J-\phi_S)F_{UT}^{\sin(3\phi_J-\phi_S)}\nonumber\\
&\left.\quad\quad\quad\quad\quad\quad\quad\quad\quad +\sqrt{2\epsilon(1+\epsilon)}\sin\phi_S F_{UT}^{\sin\phi_S}+\sqrt{2\epsilon(1+\epsilon)}\sin(2\phi_J-\phi_S)F_{UT}^{\sin(2\phi_J-\phi_S)}\right]\nonumber\\
&\quad\quad\quad\quad\quad\quad\quad\quad\quad +|{\bm S}_{\perp}|\lambda_e\left[\sqrt{1-\epsilon^2}\cos(\phi_J-\phi_S)F_{LT}^{\cos(\phi_J-\phi_S)}
+\sqrt{2\epsilon(1-\epsilon)}\cos\phi_S F_{LT}^{\cos\phi_S}\right.\nonumber\\
&\quad\quad\quad\quad\quad\quad\quad\quad\quad \left.\left.+\sqrt{2\epsilon(1-\epsilon)}\cos(2\phi_J-\phi_S)F_{LT}^{\cos(2\phi_J-\phi_S)}\right]\right\}\,,\label{eq:diff_cross_section}
\end{align}
where $\alpha$ is the fine structure constant and
\begin{align}
\epsilon=\frac{1-y-\frac{1}{4}\gamma^2 y^2}{1-y+\frac{1}{2}y^2+\frac{1}{4}\gamma^2y^2}\,.
\end{align}
The structure functions $F$'s on the right-hand side of Eq.~(\ref{eq:diff_cross_section}) depend
on $x, Q^2, z$, and $P_{J\perp}^2$. The subscripts $A,B,C$ of $F_{AB,C}$ denote the polarizations of the incoming lepton, the incoming nucleon, and the virtual photon respectively, with $U$ standing for unpolarized, $T$ for transversely polarized, and $L$ for longitudinally polarized. 
Up to corrections of $\mathcal{O}(M^2/Q^2)$, we have $\psi\approx \phi_S$, $\gamma^2\approx 0$, and the following approximations for the
coefficients of the $F$'s in Eq.~(\ref{eq:diff_cross_section}):
\begin{align}
&\frac{y^2}{2(1-\epsilon)}\approx 1-y+\frac{1}{2}y^2,\nonumber\\
&\frac{y^2}{2(1-\epsilon)}\epsilon \approx 1-y,\nonumber\\
&\frac{y^2}{2(1-\epsilon)}\sqrt{2\epsilon(1+\epsilon)} \approx (2-y)\sqrt{1-y},\nonumber\\ 
& \frac{y^2}{2(1-\epsilon)}\sqrt{2\epsilon(1-\epsilon)} \approx y\sqrt{1-y}\,,\nonumber\\
& \frac{y^2}{2(1-\epsilon)}\sqrt{1-\epsilon^2} \approx y\left(1-\frac{1}{2}y\right)\,.
\end{align} 
The structure functions $F$'s are convolutions of the nucleon TMD PDFs and the jet functions. As noted in Eq.~(\ref{eq:jet_expd}), there are two
jet functions $\mathcal{J}_1(z,k_T^2)$ and $\mathcal{J}_T(z,k_T^2)$ at leading power in $\Lambda_{QCD}/Q$. At leading order in $M/Q$, there are eight
quark TMD PDFs for the nucleon~\cite{Angeles-Martinez:2015sea}, each encoding a specific correlation between the quark spin and the proton spin as depicted in Table~\ref{tb:TMD_PDF}. Among these eight TMD PDFs, $f_1$, $g_{1L}$, $f^\perp_{1T}$, and $g_{1T}$ are chiral-even, while 
$h_1^\perp$, $h_{1L}^\perp$, $h_{1T}$, and $h_{1T}^\perp$ are chiral-odd.
For any functions 
$w({\bm p}_T,{\bm k}_T)$, $f(x,p_T^2)$, and $\eta(z,k_T^2)$, we define
\begin{align}
{\cal C}[w f \eta]\equiv x\sum_a e_a^2\int d^2\bm{p}_T\int d^2\bm{k}_T \,
\delta^{(2)}\left(\bm{p}_T-\bm{k}_T-\bm{P}_{J\perp}/z\right)
w(\bm{p}_T,\bm{k}_T)f^a(x,p_T^2)\eta^a(z,k_T^2)\,,
\end{align}
where $a$ denotes a quark or antiquark flavor.
At leading order in $\alpha_s$ and $M/Q$, the nonvanishing $F$'s are given by
\begin{align}
& F_{UU,T}={\cal C}[f_1{\cal J}_1]\,,\label{eq:F_UU,T}\\
& F_{LL}={\cal C}[g_{1L}{\cal J}_1]\,,\\
& F^{\sin(\phi_J-\phi_S)}_{UT,T} ={\cal C}\left[-\frac{\hat{\bm{h}}\cdot \bm{p}_T}{M}
f^\perp_{1T}{\cal J}_1\right]\,,\\
& F^{\cos(\phi_J-\phi_S)}_{LT}={\cal C}\left[\frac{\hat{\bm{h}}\cdot \bm{p}_T}{M}
g_{1T}{\cal J}_1\right]\,,\\
& F_{UU}^{\cos(2\phi_J)} ={\cal C}\left[-\frac{(2(\hat{\bm{h}}\cdot \bm{k}_T)(\hat{\bm{h}}\cdot \bm{p}_T)-\bm{k}_T\cdot\bm{p}_T)}{M}
h_1^\perp{\cal J}_T\right]\,,\\
& F_{UL}^{\sin(2\phi_J)} ={\cal C}\left[-\frac{(2(\hat{\bm{h}}\cdot \bm{k}_T)(\hat{\bm{h}}\cdot \bm{p}_T)-\bm{k}_T\cdot\bm{p}_T)}{M}
h_{1L}^\perp{\cal J}_T\right]\,,\\
& F^{\sin(\phi_J+\phi_S)}_{UT} ={\cal C}\left[-\hat{\bm{h}}\cdot \bm{k}_T
h_1{\cal J}_T\right]\,,\\
& F_{UT}^{\sin(3\phi_J-\phi_S)} ={\cal C}\left[\frac{2(\hat{\bm{h}}\cdot \bm{p}_T)(\bm{p}_T\cdot \bm{k}_T)+
\bm{p}_T^2(\hat{\bm{h}}\cdot \bm{k}_T)-4(\hat{\bm{h}}\cdot \bm{p}_T)^2(\hat{\bm{h}}\cdot \bm{k}_T)}{2M^2}
h_{1T}^\perp{\cal J}_T\right]\,,\label{eq:F_UT_sin3}
\end{align}
where $\hat{\bm{h}}=\bm{P}_{J\perp}/|\bm{P}_{J\perp}|$ and $h_1= h_{1T}+\frac{\bm{p}_T^2}{2M^2}h_{1T}^\perp$\,.
From Eqs.~(\ref{eq:F_UU,T})-(\ref{eq:F_UT_sin3}), we see that the T-even jet function $\mathcal{J}_1$ couples to the chiral-even 
TMD PDFs, while the T-odd jet function $\mathcal{J}_T$ couples to the chiral-odd TMD PDFs. With both the T-even and T-odd jet functions, one can thus access all eight TMD PDFs at leading twist. 

\begin{table}[htbp]
\begin{center}
\begin{tabular}{|c|c|c|c|}
\hline
\backslashbox{hadron}{quark} & unpolarized & chiral & transverse \\
\hline
$U$& $f_1$ &  & $h_1^\perp$ \\
\hline
$L$&  & $g_{1L}$ & $h^\perp_{1L}$ \\
\hline
$T$& $f^\perp_{1T}$ & $g_{1T}$ & $h_{1T},h^{\perp}_{1T}$\\
\hline
\end{tabular}
\caption{The eight TMD PDFs of a nucleon at leading twist.}
\end{center}
\label{tb:TMD_PDF}
\end{table}

With the known proton TMD PDFs and partial knowledge on the jet functions, 
we can make preliminary predictions on the azimuthal asymmetries associated with the jet-axis probe in DIS machines such as the EIC, the EicC, and HERA. 
As notes in Section~\ref{sec:T_odd_jet}, the $z$-dependence of the jet functions become trivial with the WTA jet-axis definition. In the following,
we will adopt the spherically-invariant jet algorithm Eq.~(\ref{eq:spherical_jet_alg}) with $R=1$ and the WTA scheme, so that 
\begin{align}
{\cal J}(z,k_T,R)&=\delta(1-z) {\mathfrak J}(k_T)+ {\cal O}\left( \frac{k_T^2}{E_J^2R^2} \right)\,, \label{eq:WTA_EJ}
\end{align}
where 
\begin{align}
{\mathfrak J}(k_T)&={J}(k_T^2)\frac{\slashed{\bar{n}}}{2}+i{J}_T(k_T^2)\frac{\slashed{k}_T\slashed{\bar{n}}}{2}\,.
\end{align}
We will study the azimuthal asymmetries associated with the terms $|{\bm S}_\perp|\epsilon\sin(\phi_J+\phi_S)F_{UT}^{\sin(\phi_J+\phi_S)}$ 
and $\epsilon\cos(2\phi_J)F_{UU}^{\cos 2\phi_J}$ in Eq.~(\ref{eq:diff_cross_section}). These terms probe the transversity $h_1$ and the Boer-Mulders function $h_1^\perp$. These terms
can be singled out by specific modulated cross sections. We define a $|{\bm P}_{J\perp}|$-distribution of asymmetry that probes the transversity by
\begin{align}
A^{\sin(\phi_J+\phi_S)}(|{\bm P}_{J\perp}|)&=\frac{2}{|\bm{S}_\perp|\int d\sigma \epsilon}\int d\sigma\sin(\phi_J+\phi_S)\label{eq:A_transversity_1} \\
&=\frac{\langle \epsilon F_{UT}^{\sin(\phi_J+\phi_S)}\rangle}{\bar{\epsilon} \langle F_{UU,T}\rangle}\,,\label{eq:A_transversity_2}
\end{align}
where by $\langle X \rangle$ we mean
\begin{align}
\langle X \rangle&=\int dx \int dy \int d\phi_S \int dz \int d\phi_J\, \frac{\alpha}{xyQ^2}\frac{y^2}{2(1-\epsilon)}X\,,
\end{align}
and 
\begin{align}
\bar{\epsilon}=\frac{\int d\sigma \epsilon}{\int d\sigma}\,.
\end{align}
We can write $F_{UU,T}$ and $F_{UT}^{\sin(\phi_J+\phi_S)}$ as
\begin{align}
F_{UU,T}&=x\sum_a e_a^2\int \frac{d^2b}{(2\pi)^2}e^{-i{\bm P}_{J\perp}\cdot{\bm b}}\tilde{f}^a_1(x,b^2)\tilde{J}^a(b^2)\,,\label{eq:F_UU}\\
F_{UT}^{\sin(\phi_J+\phi_S)}&=-ix\sum_a e_a^2\int \frac{d^2b}{(2\pi)^2}\frac{P_{J\perp}^i}{|\bm{P}_{J\perp}|}
e^{-i{\bm P}_{J\perp}\cdot{\bm b}}\tilde{h}^a_1(x,b^2)\partial_{b^i}\tilde{J}^a_T(b^2)\,,\label{eq:F_UT}
\end{align}
where we have used the Fourier transforms $\tilde{f}_1(x,b^2)=\int d^2p_T\, e^{-i{\bm p}_T\cdot{\bm b}}f_1(x,p_T^2)$ and
$\tilde{J}(b^2)=\int d^2k_T\, e^{-i{\bm k}_T\cdot{\bm b}}J(k_T^2)$, and similarly for $\tilde{h}_1$ and $\tilde{J}_T$. 
Similar to the treatment in Ref.~\cite{Kang:2014zza}, we include the effect of evolution by including a 
Sudakov factor in $b$-space 
\begin{align}
\tilde{f}_1(x,b^2,Q)=e^{-S_{\rm pert}-S_{\rm NP}}\tilde{f}_1(x,b^2,Q_0)\,,\\
\tilde{J}(b^2,Q)=e^{-S_{\rm pert}-S_{\rm NP}}\tilde{J}(b^2,Q_0)\,,
\end{align}
and similarly for $\tilde{h}_1$ and $\tilde{J}_T$, where 
\begin{align}
S_{\rm pert}&=\int^Q_{\mu_b} \frac{d\mu}{\mu}\,\frac{\alpha_s(\mu)}{2\pi}C_F
\ln \frac{Q^2}{\mu^2}\,,\\
S_{\rm NP}&=\frac{g_2}{2}\ln\left(\frac{b}{b_*}\right)\ln\left(\frac{Q}{Q_0}\right)\,,
\end{align}
with $\mu_b=1.22/b_*$, $b_*=b/\sqrt{1+b^2/b_*^2}$, $Q_0=1.549$~GeV, and $g_2=0.84$. Here,
we have adopted the expression of $S_{\rm pert}$ at leading logarithm.

\begin{figure}[htbp]
  \begin{center}
 \subfigure[]{ 
   \includegraphics[scale=0.3]{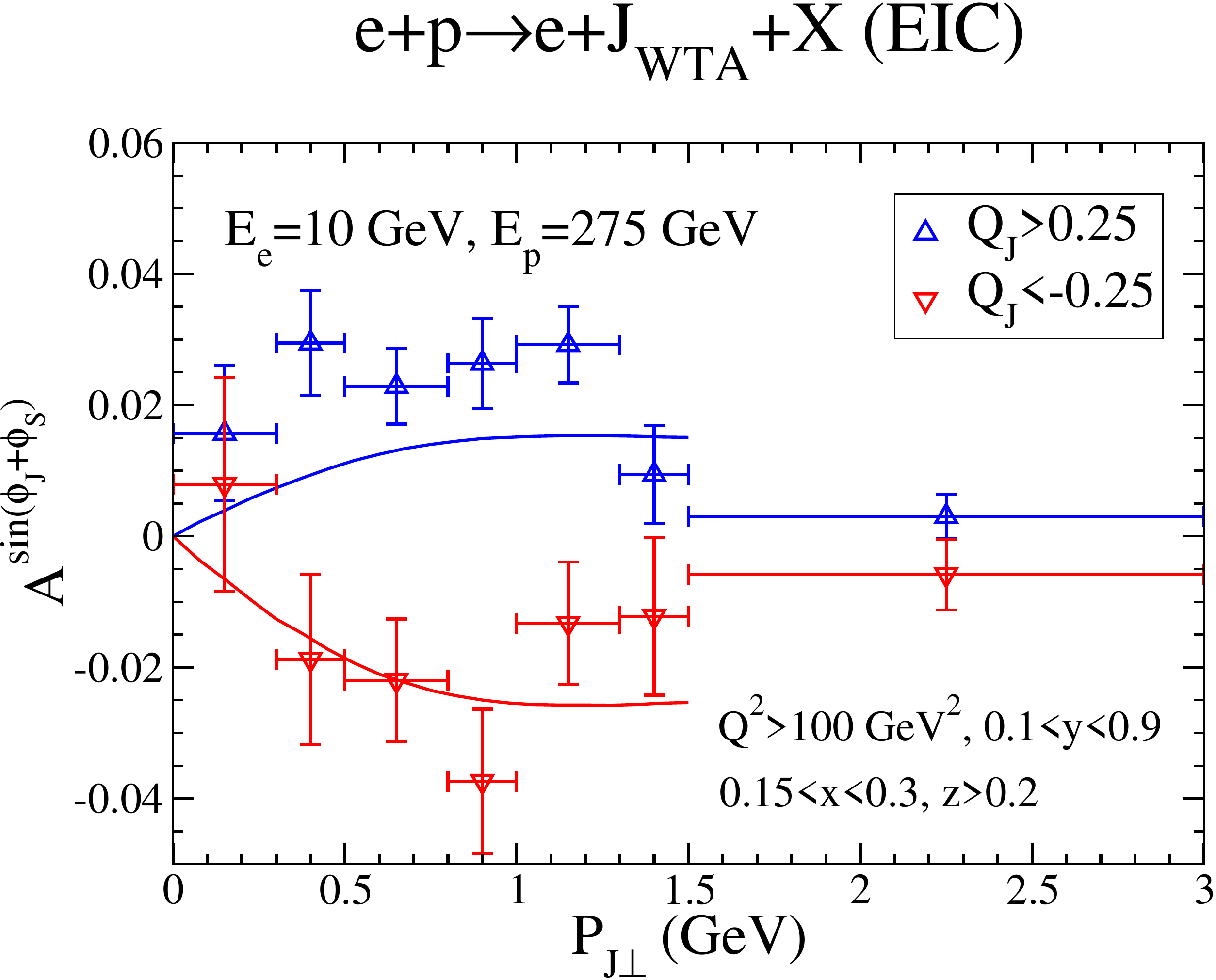}
     }
   \subfigure[]{ 
   \includegraphics[scale=0.3]{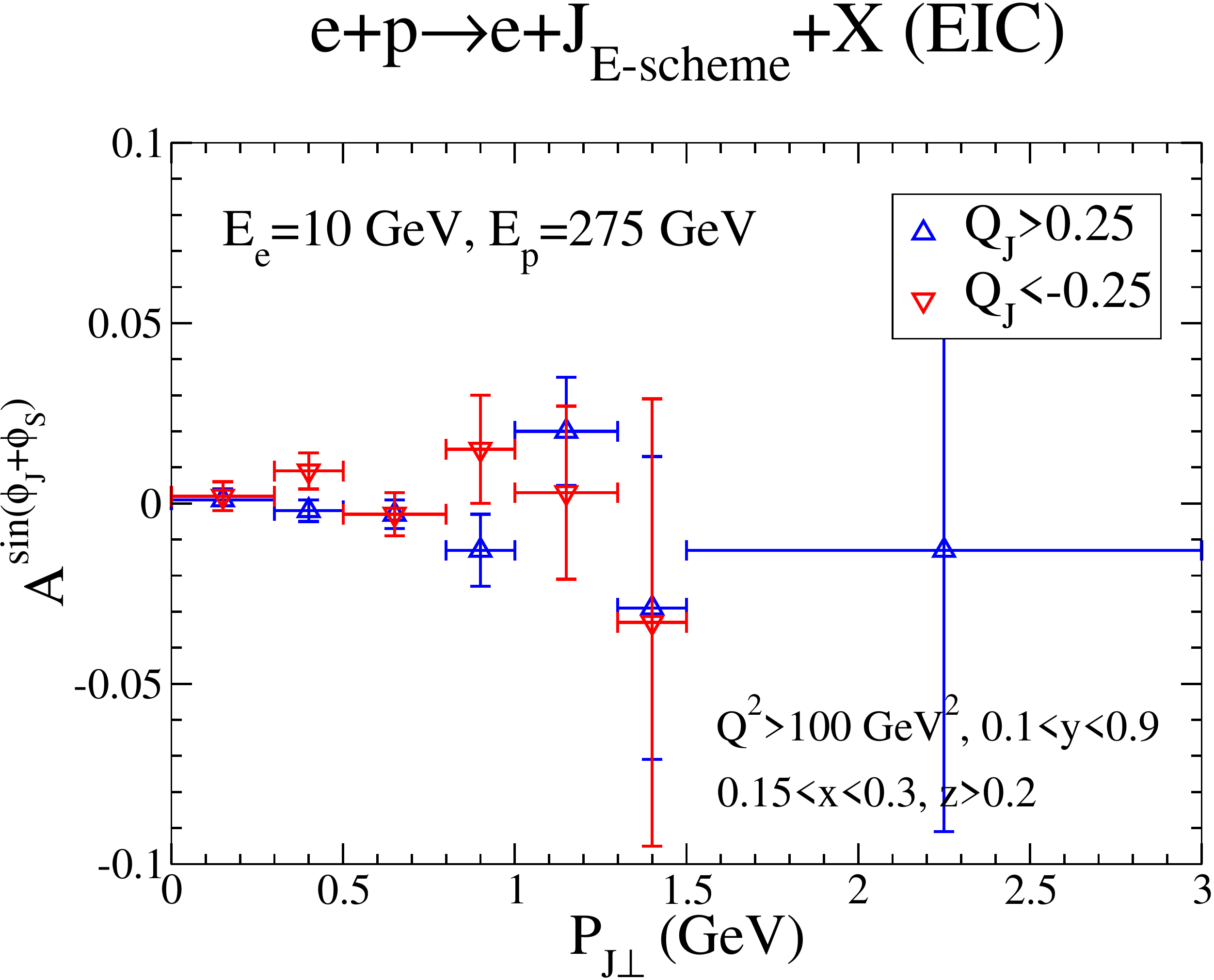}
  }
\caption{Azimuthal asymmetry $A^{\sin(\phi_J+\phi_S)}$ in the jet-axis probe at the EIC in (a) the WTA scheme and (b) the E-scheme.
The data points with error bars are from event generator simulations. The solid lines in (a)
are from Eq.~(\ref{eq:A_transversity_2}). }
  \label{fg:asym_transversity_EIC}
 \end{center}
\end{figure}

For the transversity, we use the fitted parametrized form Ref.~\cite{Martin:2014wua}. For the jet functions, 
although not mandatory, we will apply the jet charge measurement in order to enhance flavor separation. This amounts to
replacing the overall normalizations of the jet functions by the charge bins $r_a$, whose values for ${J}^a$
and for jet charge $Q_J>0.25$ and $Q_J<-0.25$ have been obtained in Ref.~\cite{Kang:2020fka} and will be used in this work.
For the charge bins associated with $J^a_T$, we will take them as a product $N_a r_a$, where $N_a$ is the ratio
of the overall normalization of the pion Collins function $H_1^{\perp a}$ to the overall normalization of fragmentation function $D_1^a$ as 
obtained in Ref.~\cite{deFlorian:2007aj}.
For the $p_T^2$-dependence of $\tilde{J}$ and $\tilde{J}_T$, we use that of the 
fragmentation function and the Collins function of the pion as obtained in Ref.~\cite{deFlorian:2007aj}. Figure~\ref{fg:asym_transversity_EIC} (a) 
shows the predictions of the asymmetry 
$A^{\sin(\phi_J+\phi_S)}(|{\bm P}_{J\perp}|)$ at the EIC according to Eq.~(\ref{eq:A_transversity_2}) (solid lines) and 
Eq.~(\ref{eq:A_transversity_1}) (data points)
from simulations using \textsc{Pythia} 8.2~\cite{Sjostrand:2014zea} with the package \textsc{StringSpinner}~\cite{Kerbizi:2021pzn}, which incorporates spin interactions in the event generator.
From Fig.~\ref{fg:asym_transversity_EIC} (a), we see that the theoretical predictions on the $A^{\sin(\phi_J+\phi_S)}(|{\bm P}_{J\perp}|)$ distribution from the factorization formula Eq.~(\ref{eq:diff_cross_section}) roughly agree with the event generator simulations.
In Figure~\ref{fg:asym_transversity_EIC} (b), we show the prediction of $A^{\sin(\phi_J+\phi_S)}(|{\bm P}_{J\perp}|)$ with the E-scheme for the jet-axis definition from \textsc{Pythia} 8.2~\cite{Sjostrand:2014zea} with \textsc{StringSpinner}, with the same kinematic setting as 
Fig.~\ref{fg:asym_transversity_EIC} (a). We see that the asymmetry no longer exists in the E-scheme. This is because the asymmetry is
nonvanishing only when the direction of the fragmenting parton which initiates the jet differs with that of the jet axis, which 
hardly the case in the E-scheme. In this sense, by choosing different jet axes we are able to ``film" the nonperturbative dynamics of QCD.

\begin{figure}[htbp]
  \begin{center}
   \includegraphics[scale=0.3]{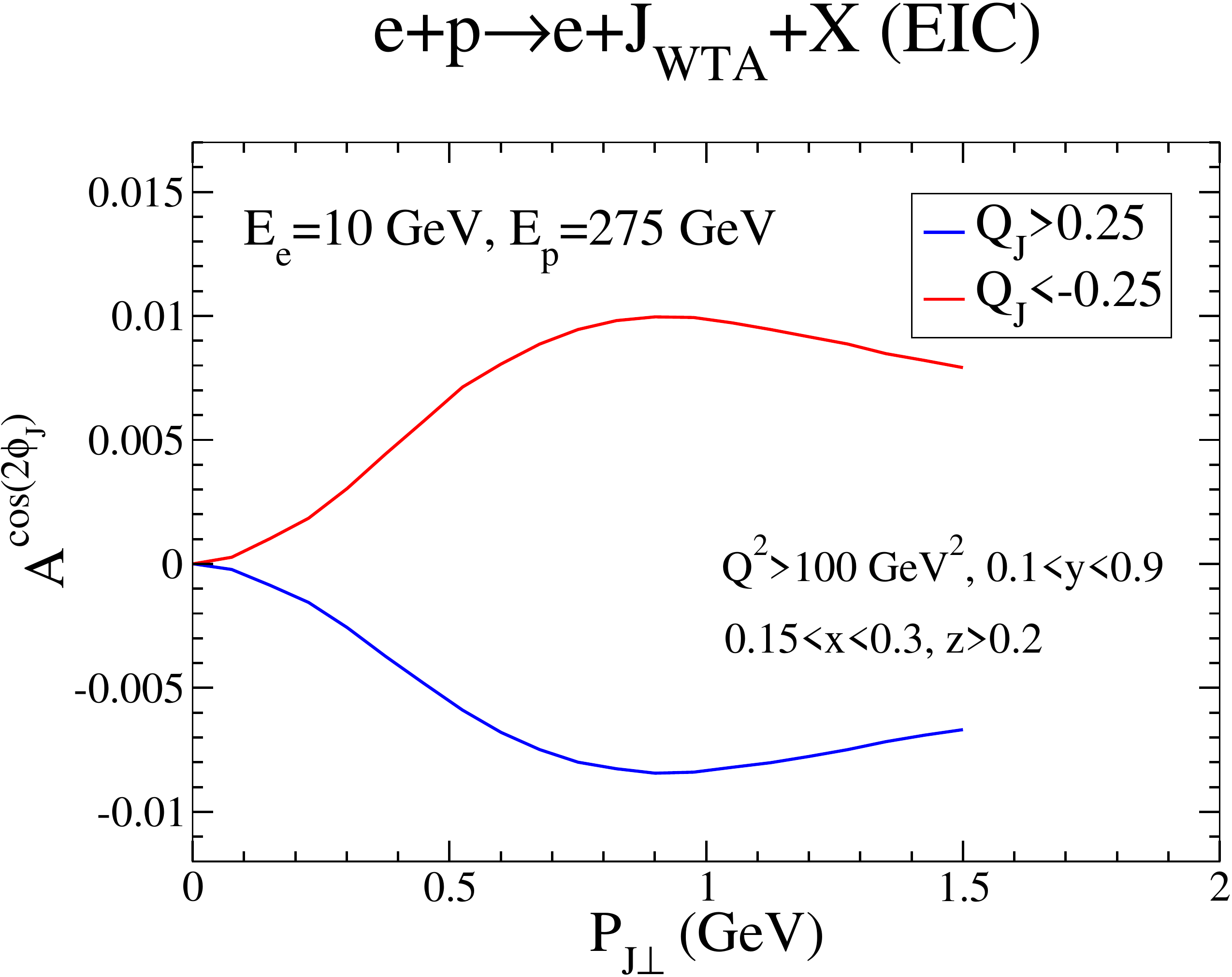} 
   \caption{Azimuthal asymmetry $A^{\cos(2\phi_J)}$ in the jet-axis probe at the EIC as predicted from Eq.~(\ref{eq:A_Boer_2}). }
   \label{fg:asym_Boer_EIC}
 \end{center}
\end{figure}

Likewise, we can make predictions on the asymmetry that probes the Boer-Mulders function defined by
\begin{align}
A^{\cos(2\phi_J)}(|{\bm P}_{J\perp}|)&=\frac{2}{\int d\sigma \epsilon}\int d\sigma\cos(2\phi_J)\label{eq:A_Boer_1}\\
&=\frac{\langle \epsilon F_{UU}^{\cos(2\phi_J)}\rangle}{\bar{\epsilon} \langle F_{UU,T}\rangle}\,.\label{eq:A_Boer_2}
\end{align}
The structure function $F_{UU}^{\cos(2\phi_J)}$ can be written as
\begin{align}
F_{UU}^{\cos(2\phi_J)}&=-\frac{x}{M}\sum_a e_a^2\int \frac{d^2b}{(2\pi)^2}e^{-i{\bm P}_{J\perp}\cdot{\bm b}}
\left[\frac{2}{|{\bm P}_{J\perp}|^2}\left(P^i_{J\perp}\cdot\partial_i\tilde{h}^{\perp a}_1\right)\left(P^j_{J\perp}\cdot\partial_j\tilde{J}^{a}_T\right)-
\partial_i\tilde{h}^{\perp a}_1\partial_i\tilde{J}_T\right]
\,,\label{eq:F_UU_cos2}
\end{align}
where $\tilde{h}_1^\perp$ and $\tilde{J}_T$ are the Fourier transforms of $h_1^\perp$ and $J_T$ respectively. We adopt the Boer-Mulders
functions obtained from Ref.~\cite{Barone:2009hw}. The predictions on $A^{\cos(2\phi_J)}(|{\bm P}_{J\perp}|)$ at the EIC according to Eq.~(\ref{eq:A_Boer_2})
are shown in Fig.~(\ref{fg:asym_Boer_EIC}).

\begin{figure}[htbp]
  \begin{center}
 \subfigure[]{ 
   \includegraphics[scale=0.3]{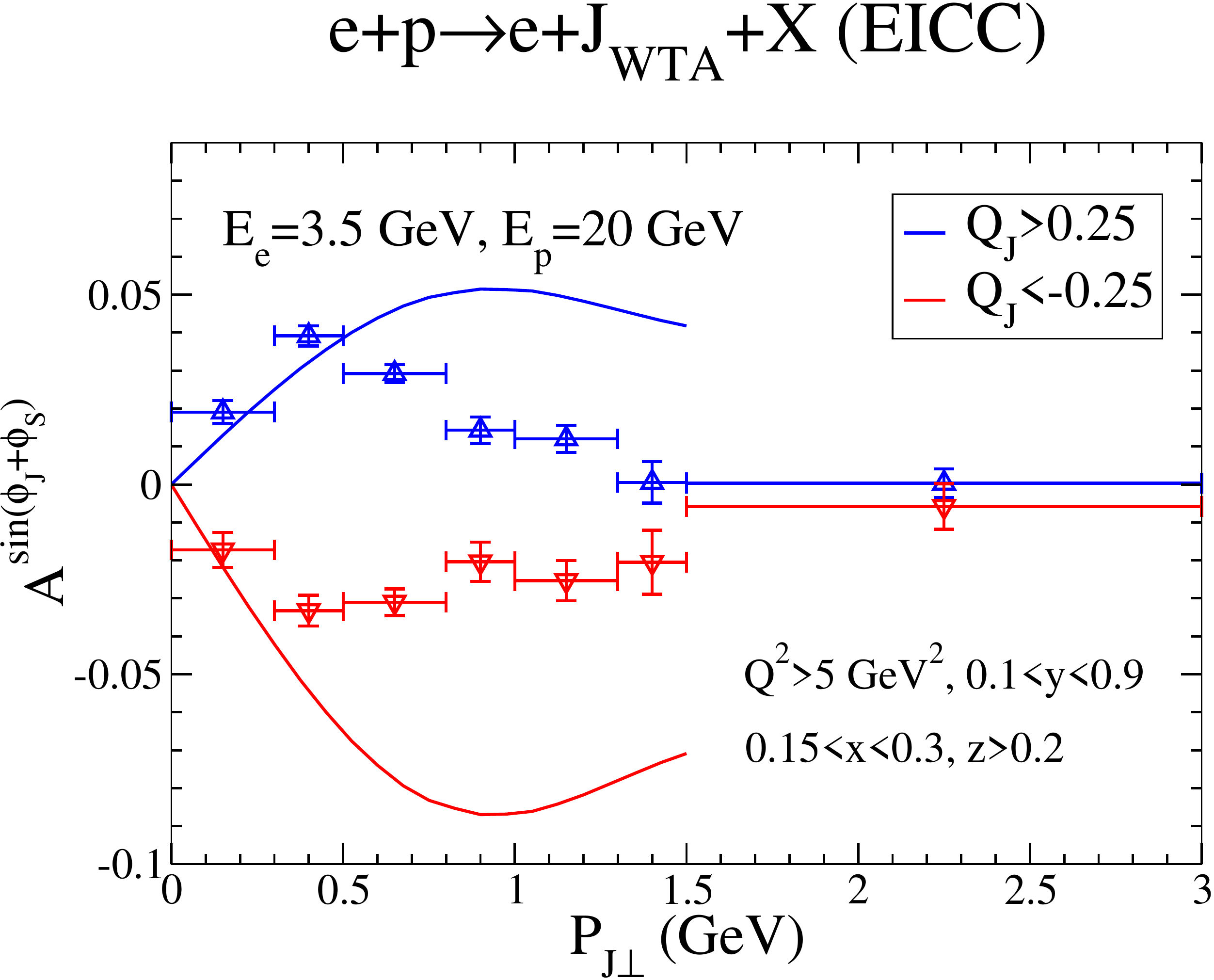}
     }
   \subfigure[]{ 
   \includegraphics[scale=0.3]{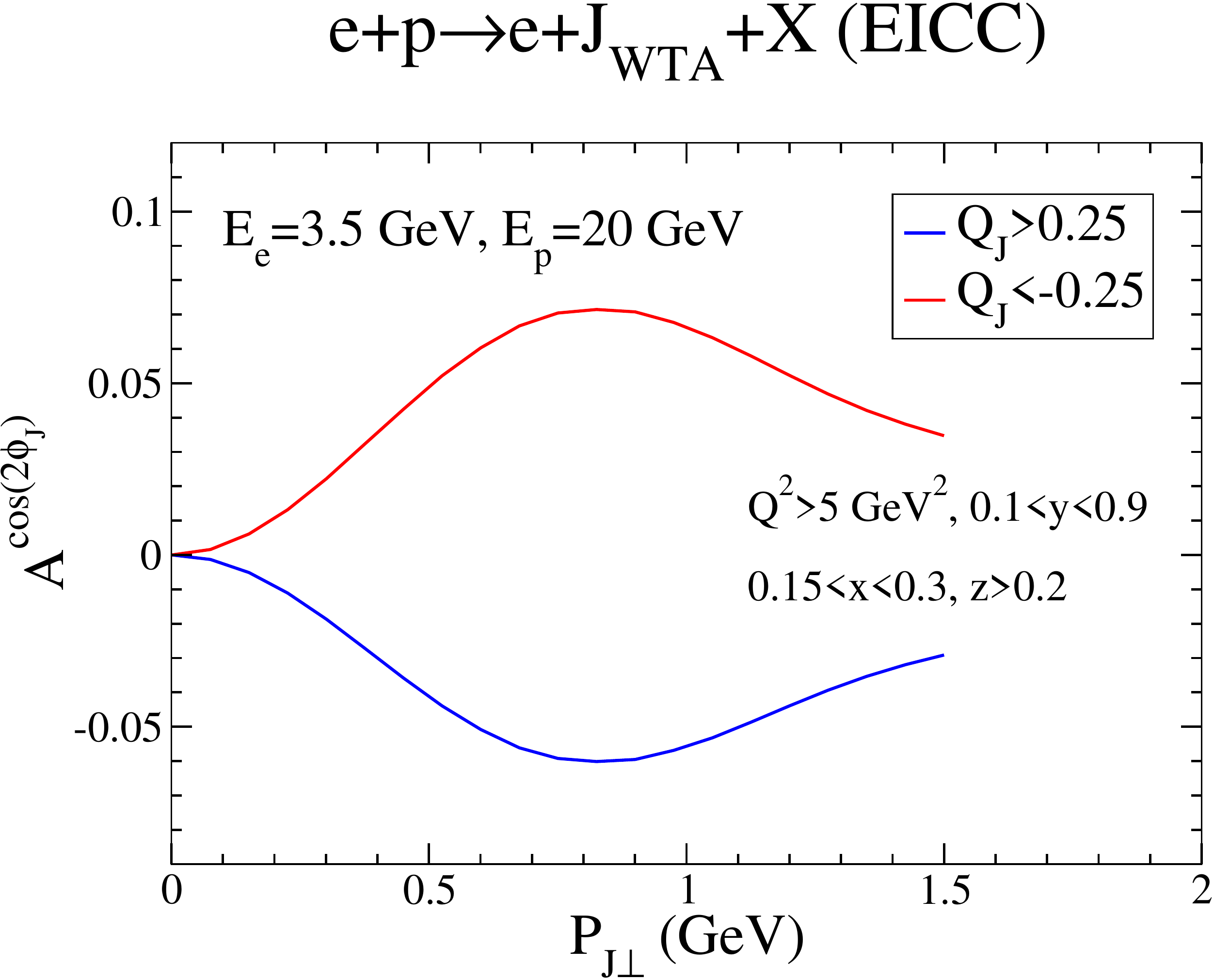}
  }
\caption{Azimuthal asymmetries $A^{\sin(\phi_J+\phi_S)}$ (a) and $A^{\cos(2\phi_J)}$ (b) in the jet-axis probe at the EicC. The data points with error bars are from event generator 
simulations and the lines are from the factorization formulae. }
  \label{fg:asym_EICC}
 \end{center}
\end{figure}

The predictions of $A^{\sin(\phi_J+\phi_S)}$ and $A^{\cos(2\phi_J)}$ at the EicC are shown in Fig.~\ref{fg:asym_EICC}. As in Fig.~\ref{fg:asym_transversity_EIC} (a) and Fig.~\ref{fg:asym_Boer_EIC}, the data points with error bars are from event generator 
simulations and the lines are from the factorization formulae.

\begin{figure}[htbp]
  \begin{center}
 \subfigure[]{ 
   \includegraphics[scale=0.3]{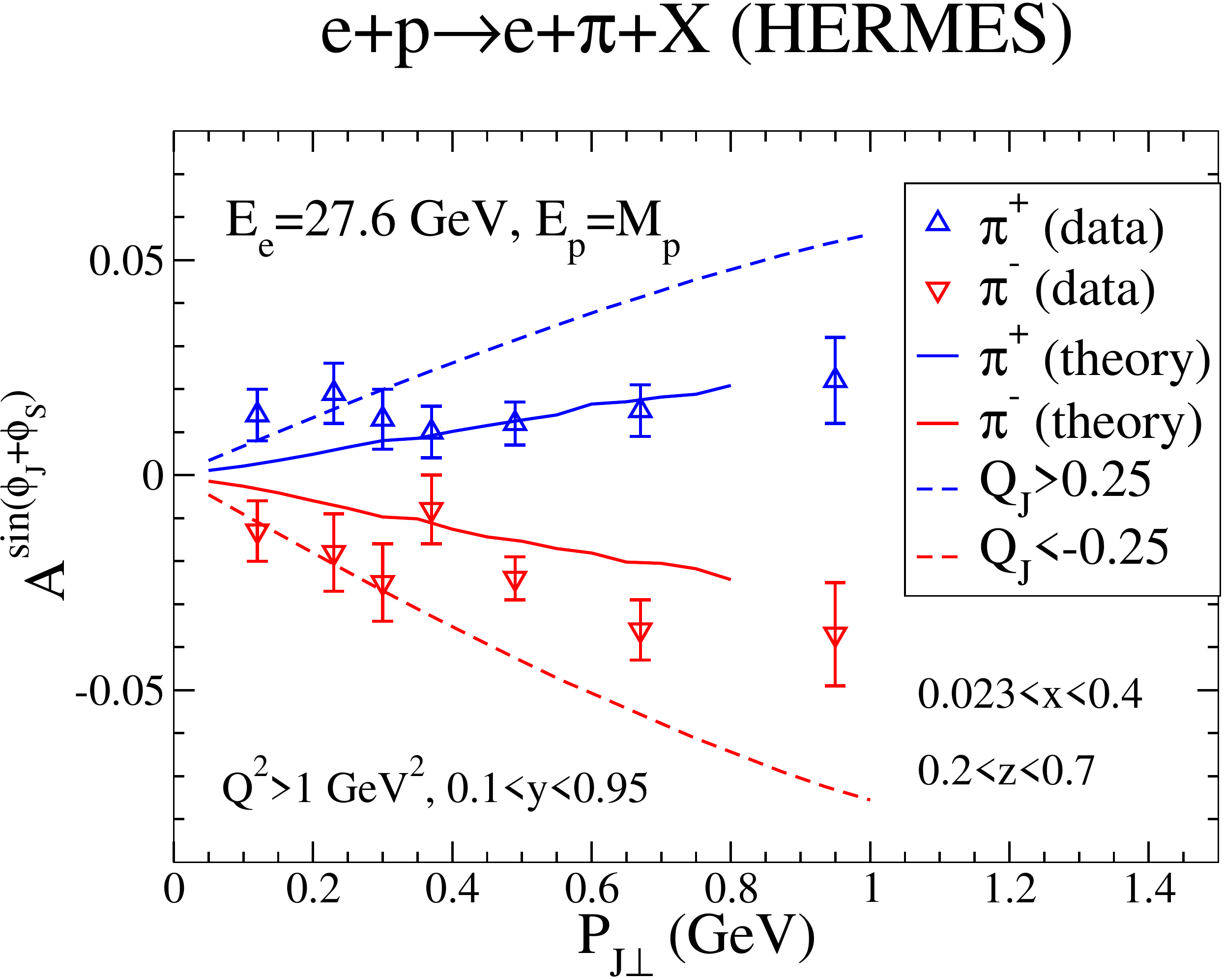}
     }
   \subfigure[]{ 
   \includegraphics[scale=0.3]{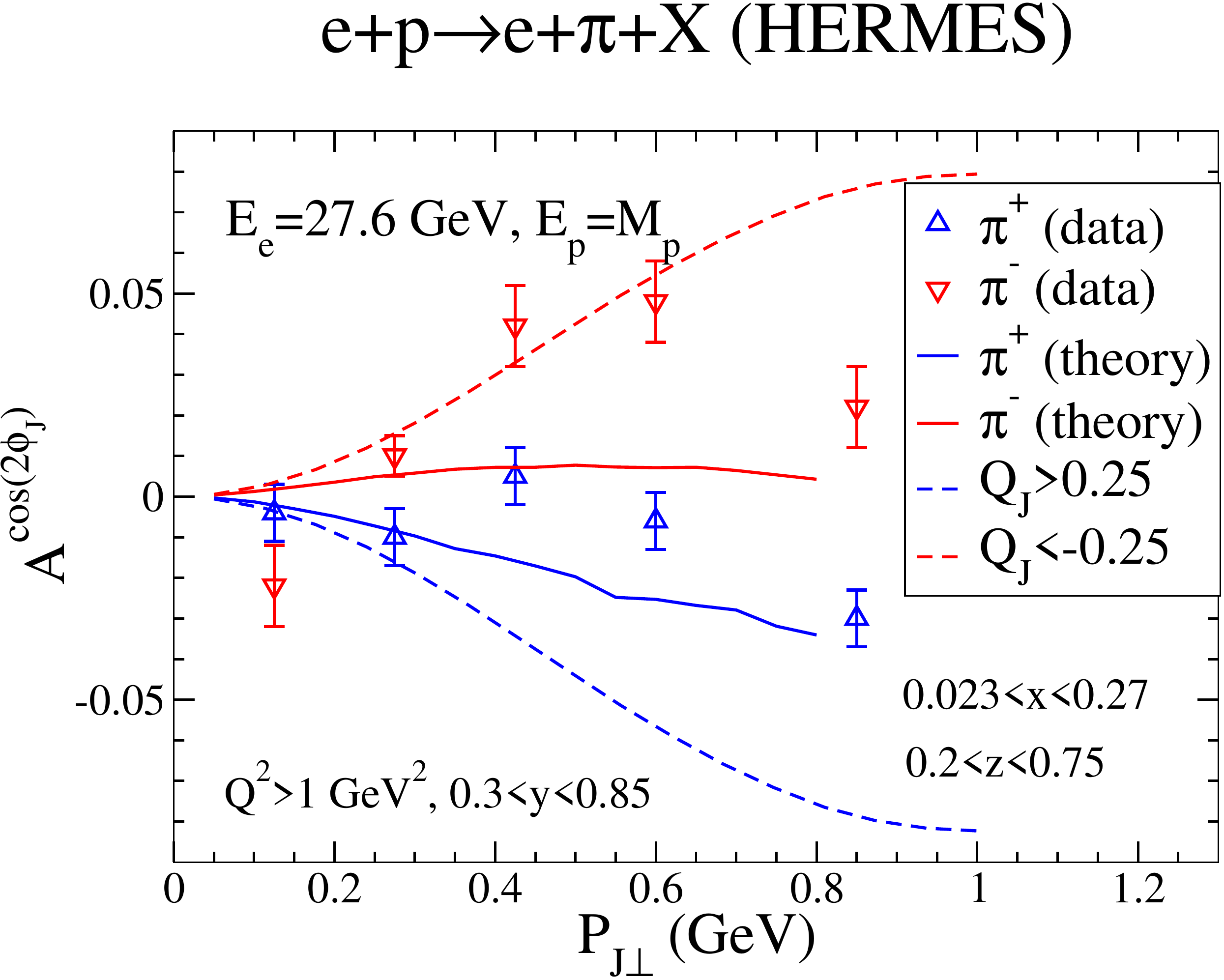}
  }
\caption{Azimuthal asymmetries $A^{\sin(\phi_J+\phi_S)}$ and $A^{\cos(2\phi_J)}$ at HERA. The data points with error bars are from the HERMES experiment for pion production. The solid lines are predictions from the factorization formulae for pion production. The dashed line are predictions from the factorization formulae for the jet-axis probe. }
  \label{fg:asym_Hermes}
 \end{center}
\end{figure}

For the sake of comparison with data in SIDIS, in Fig.~\ref{fg:asym_Hermes} we show the asymmetries $A^{\sin(\phi_J+\phi_S)}(|{\bm P}_{J\perp}|)$
(a) and $A^{\cos(2\phi_J)}(|{\bm P}_{J\perp}|)$ (b) at HERA with predictions for jets from Eqs.~(\ref{eq:A_transversity_2}) and~(\ref{eq:A_Boer_2}) 
(dashed lines), predictions for pion production from the parallels of Eqs.~(\ref{eq:A_transversity_2}) and~(\ref{eq:A_Boer_2}) as 
in~Refs.~\cite{Bacchetta:2006tn,Barone:2009hw} (solid lines), and data points for pion production from the HERMES experiment~\cite{HERMES:2010mmo,HERMES:2012kpt}
(data points with error bars). From Fig.~\ref{fg:asym_Hermes}, we see that the T-odd jet function 
does give azimuthal asymmetries with sizes and shapes similar to those in SIDIS, and so should be observable 
even at low-energy machines.

\section{back-to-back dijet production in $e^+e^-$ annihilation} \label{sec:ee}

\begin{figure}[htbp]
  \begin{center}
   \includegraphics[scale=0.3]{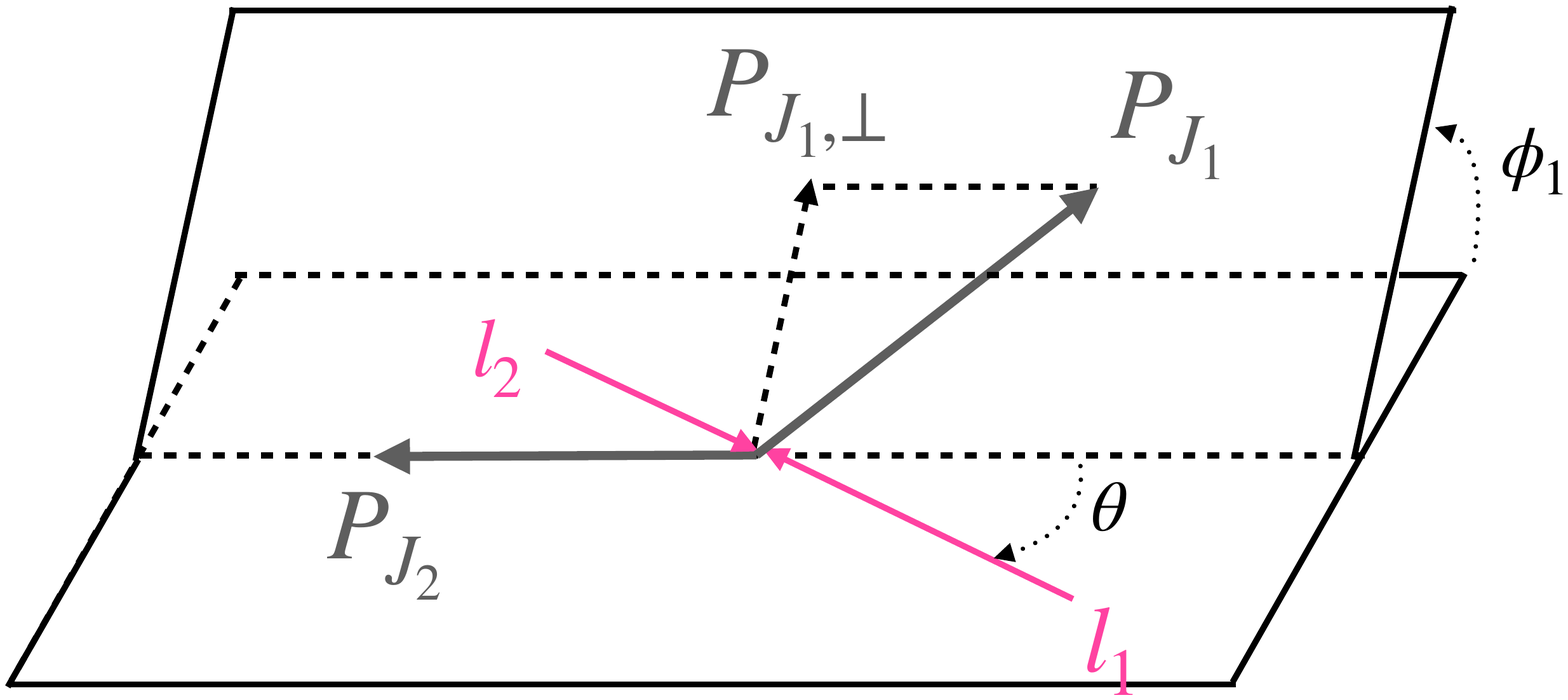} 
\caption{Back-to-back dijet production in $e^+e^-$ annihilation.}
  \label{fg:ee}
 \end{center}
\end{figure}

The T-odd jet function will give rise to novel jet phenomena in $e^+e^-$ annihilation, which are measurable at $e^+e^-$ machines.
For instance, consider back-to-back dijet production in $e^+e^-$ annihilation, as shown in Fig.~\ref{fg:ee}. We define 
${\bm q}_T=-{\bm P}_{J_1\perp}$. The back-to-back limit corresponds to $|{\bm q}_{T}|\ll \sqrt{s} R$, where $R$ is the jet radius. The azimuthal asymmetry $A$~\cite{Kang:2014zza} is given by
\begin{align}
A &=  2 \int  d\cos\theta \, \frac{ d\phi_1}{\pi} \cos(2\phi_1)A^{J_1J_2}\,,\label{eq:A}
\end{align}
where
\begin{align}
 A^{J_1J_2} &= 
 1+ \cos(2\phi_1) \frac{\sin^2\theta}{1+\cos^2\theta} \frac{F_T}{F_U}\,,   \label{eq:A_2} 
\end{align}
with
\begin{align}
F_U &= |\bm{q}_T|\, \sum_q e_q^2\, \int \frac{{d}^2b}{(2\pi)^2} e^{i\bm{q}_T\cdot\bm{b}} \tilde{J}_1^q(b^2){ \tilde{J}}^{\bar q}_1(b^2) \,, 
\label{eq:F_U}\\
F_T &=|\bm{q}_T| \, \sum_q \, e_q^2\, \int \frac{{d}^2b}{(2\pi)^2}e^{i\bm{q}_T\cdot\bm{b}}
\left( 2 \frac{q_T^i q_T^j}{|\bm{q}_T|^2} 
- \delta^{ij}  \right) \partial_{b^i} \tilde{J}^q_{T}(b^2) \partial_{b^j}\tilde{J}^{\bar q}_{T}(b^2) \,.\label{eq:F_T}
\end{align} 

\begin{figure}[htbp]
  \begin{center}
   \includegraphics[scale=0.4]{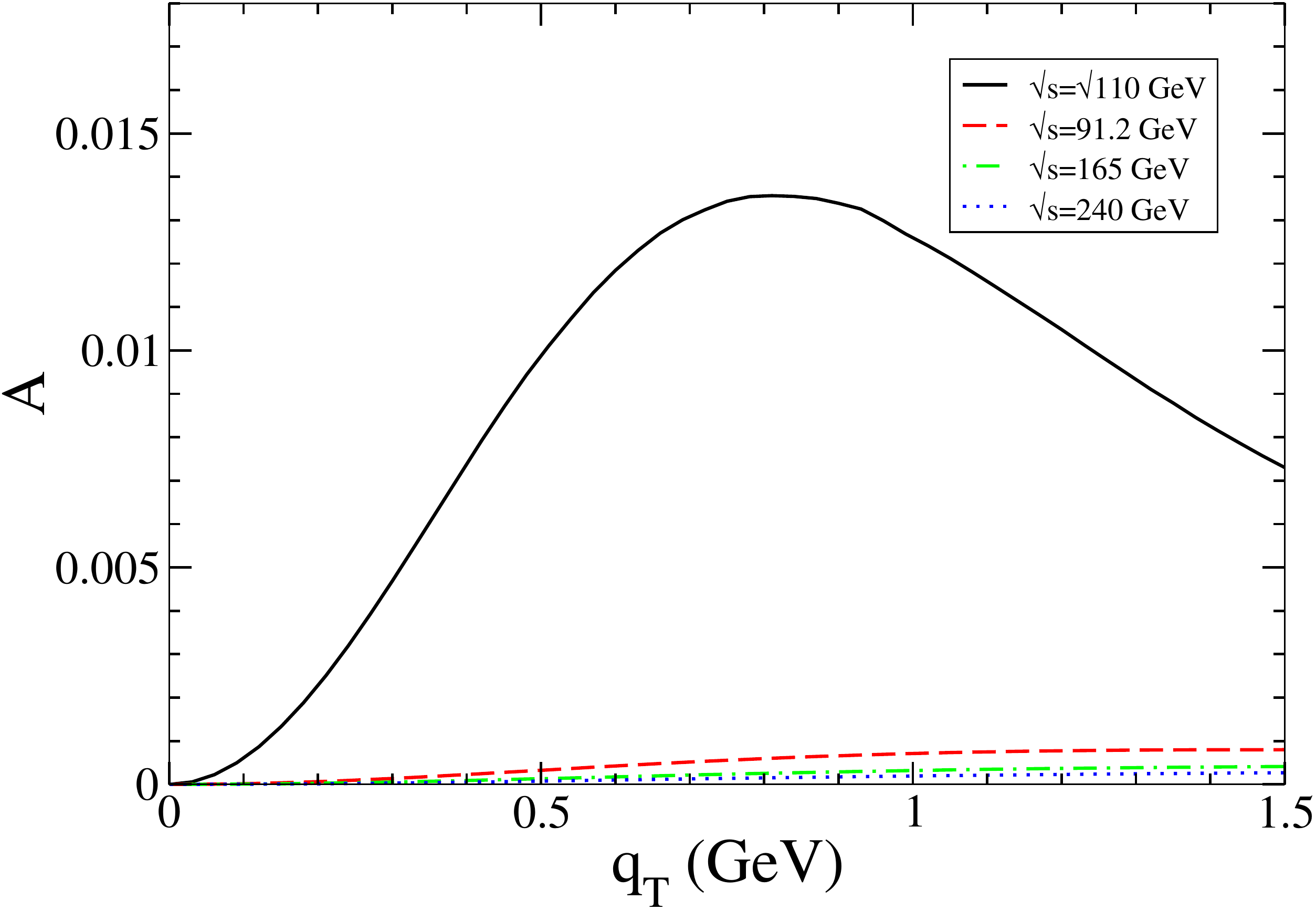} 
\caption{Azimuthal asymmetry $A$ for dijet production in $e^+e^-$ annihilation as predicted by Eq.~(\ref{eq:A_2}) at Belle, LEP and the CEPC, with $Q_J>0.25$ for one of the jets and $Q_J<-0.25$ for the other.}
 \label{fg:R_plot}
  \end{center}
\end{figure}

In Fig.~\ref{fg:R_plot}, we plot the asymmetry $A$ as a function of $|\bm{q}_T|$ as predicted by 
Eq.~(\ref{eq:A_2}) for four different values of $\sqrt{s}$. To enhance the sensitivity, we have demanded that $Q_J>0.25$ for one of the jets and $Q_J<-0.25$ for the other.
The value $\sqrt{s}=\sqrt{110}$~GeV corresponds to the Belle experiment. The values $\sqrt{s}=91.2$~GeV, $165$~GeV, and $240$~GeV correspond to the $Z$-threshold, the $W$-threshold, and the $Z$-Higgs-threshold respectively at LEP as well as the CEPC. One can see that the asymmetry is more significant at low-energy machines.

\section{Summary and outlook} \label{sec:summary}

In this work, we reinterpreted the jet clustering procedure as a way to define an axis, which together with the proton beam defines the transverse momentum of the vitual photon in DIS. In this way, one can use jet-axis measurements in DIS to probe the TMD PDFs of the nucleons, just like in the Drell Yan process. We provided the complete list of azimuthal asymmetries in the jet-axis probe in DIS at leading power. We showed that, by including the T-odd jet function in addition to the traditional one, all eight TMD PDFs of a nucleon at leading twist can be accessed by the jet-axis probe.  As concrete examples, within the WTA axis scheme, we demonstrated that with both event-generator simulations and predictions from the factorization formulae, 
couplings of T-odd jet function with the quark transversity and the Boer-Mulders function give rise to sizable azimuthal asymmetries at DIS machines of various energy regimes, such as the EIC, the EicC, and HERA. We also demonstrated, with event-generator simulations, how the change of the jet-axis definition induces changes in the asymmetry distributions drastically. 
We also gave predictions for the azimuthal asymmetry of back-to-back dijet production in $e^+e^-$ annihilation. The T-odd jet function has opened the door to a fully comprehensive study of nucleon 3D structure with jet probes. Further theoretical and phenomenological studies of the T-odd jet function, such as high-order corrections, evaluations of the soft function on the lattice, and fittings with experimental data, will empower the jet probe as a precision tool which is fully differential for the study of TMD physics.

\acknowledgments
W.~K.~L. and H.~X. are supported by
the National Natural Science Foundation of China under Grant No. 12022512, No. 12035007, and by the Guangdong Major Project of Basic and Applied Basic Research No. 2020B0301030008. X.~L. and M.~W. are supported by the National Natural Science Foundation of China under Grant No.~12175016.
W.~K.~L. acknowledges support by the UC Southern California Hub, with funding from the UC National Laboratories division of the University of California Office of the President.  


\bibliography{T_Odd_Pheno_Paper}

\end{document}